# USER AWARENESS MEASUREMENT THROUGH SOCIAL ENGINEERING

Tolga MATARACIOGLU[1] and Sevgi OZKAN[2]

[1]TUBITAK National Research Institute of Electronics and Cryptology (UEKAE),
Department of Information Systems Security, 06700, Ankara, TURKEY
[2]Middle East Technical University, Informatics Institute,
Department of Information Systems, 06531, Ankara, TURKEY

`matacioglu@uekae.tubitak.gov.tr, sozkan@ii.metu.edu.tr`

*ABSTRACT*

*TUBITAK National Research Institute of Electronics and Cryptology (UEKAE) Department of Information Systems Security makes social engineering attacks to Turkish public agencies within the frame of "Information Security Tests" [19]. This paper will make an analysis of the social engineering tests that have been carried out in several Turkish public agencies. The tests include phone calling to sample employees by the social engineer and trying to seize employees' sensitive information by exploiting their good faith. The aim of this research is to figure that the employees in Turkish public agencies have a lack of information security awareness and they compromise the information security principles which should be necessarily applied for any public agencies. Social engineering, both with its low cost and ability to take advantage of low technology, has taken its place in the information security literature as a very effective form of attack [8].*

## 1. INTRODUCTION

As it is well known, only a small percentage of information security is maintained by technical security measures, while its greater percentage depends on the user. Individuals in charge of information security in an organization are all of the organizational staff, with the foremost being the owner of the information and the IT personnel. About seventy percent of information theft is carried out consciously or unconsciously from within the organization [11], [12], [17]. The weakest link is analyzed to understand the level of information security. And the weakest link, in most of the cases, is unfortunately men. When an organization suffers from poor information security, the organization may face the following problems:

- The information may be collected by others.
- The organization's honor and its image in the society may get damaged (which is the worst case scenario).
- The hardware, software, data and organizational staff may suffer damages.
- Problems of inability to have timely access to important data may arise.
- Monetary losses may occur (the most insignificant part of the problem from a relative point of view).
- Time losses become inevitable.
- There may even be loss of life.

Federal Bureau of Investigation (FBI) estimates that industrial espionage cause American companies to suffer a loss amount of about one hundred billion dollars per annum [13]. [1] shows a distribution, classified by their types, of computer-related incidents that have occurred in the United States of America between the years 2001 and 2009. The graph in [1] shows that stolen laptops occupy the first place with a rate of 21%, computer hacking occupies the second place with a rate of 16%, web occupies the third place with a rate of 13%, and fraud occupies the fourth place with a rate of 8%. If we classify the attacks made using social engineering techniques under the titles of "Computer Hacking" and "Fraud", these incidents form an important portion of crimes with a rate of 24% in the USA.

[20] gives statistics about social engineering attacks performed in 2010 in the USA. Accordingly, 15 companies have been called and total of 135 conversations have been made during social engineering attacks. Approximately, 93% of the companies' information has been seized and only 8% of the employees put up resistance for sharing their sensitive information during phone calls.

Social engineering can be described as the art of gathering information that others should not disclose and share under normal conditions taking advantage of and utilizing methods of influencing and convincing. Most people think that they have a low chance of being deceived. Being aware of this shared conviction, the attacker presents his desire so cleverly that he arouses no suspicion and exploits the victim's trust [4], [6], [16], [18], [19].

There is a cliché that "The safest computer is an unplugged computer". Nobody thinks that an evil-minded person might convince somebody to go to the office and switch his computer on. Today, the perspective towards today's information security should be extended to cover such incidents too.

Some hardware such as USB memory watch, hardware keylogger, USB memory lighter, camera car key, micro SD card holder, camera pen, and camera eyeglasses that can be used in social engineering attacks, can easily be acquired with very low costs. And this seriously increases the chances of social engineering attacks in an organization.

In this research, the aim is to figure that the employees in Turkish public agencies have a lack of information security awareness and they compromise the information security principles which should be necessarily applied for any public agencies.

## 2. LITERATURE REVIEW

In this part, attack techniques being used in social engineering and protection methods have been described.

## 2.1 ATTACK TECHNIQUES

This section will cover the attack techniques employed by social engineers (white hats) or evil-minded persons (black hats) using social engineering techniques [2], [3], [5], [19], [21].

Breaching the security of an organization generally starts with the bad guy obtaining seemingly a very innocent, daily and trivial information or a document, which many persons in the organization see no reasons to protect or classify. Most social engineers will welcome the information that is seemingly harmless for an organization because such information might play a crucial role in showing themselves more convincing.

Most social engineering attacks are sophisticated procedures. However, some attackers can reach their target in a more simple and straightforward manner. Just asking you to give the information may be enough in some cases. This kind of attack is called direct attacking.

The secret to the success of social engineers is the men are too much prone to being deceived by gaining trust. This is because men can trust wrong things if they are manipulated in certain ways. A good social engineer predicts the possible questions of his victim and prepares himself for these questions.

You have a problem and there is someone wishing to help you. Can you reject that help? You will both accept and be grateful to the attacker. Never doubt that the attacker is the number one suspect of that problem.

The attacker puts himself into a pitiful position and asks for the victim's help. We always pity those in difficulty. The result: 100% success!

Software that can be downloaded free of charge! Did you ever think why you pay nothing to download them? There are also many software of this kind, but that are made for no malicious intents. Attackers take advantage of people's desire for free things or email attachments with desirable content. For example, if you are receiving an email having the wording "please click on the attached file to learn more about your increased salary" in its subject line, you'd better be careful before opening its attachment.

We tend to avoid challenges both for ourselves for others. Exploiting this positive motive, the attacker may play with the person's sense of pitying, may make him to feel guilty or may use suppression as a gun.

Conventional social engineers follow a certain route. However, the game is played on the reverse direction in some cases. This is called reverse tricking. In this method, the attacker creates a problem where the victim will be directly affected. He contacts the victim in a way by telephone and leaves his phone number, waiting the victim to call him back.

Penetration is a technique where someone from out of the organization disguises himself as a member of staff. Once in, many passwords can simply be acquired just by looking under the keyboards or on post-its on the screens. What are more worrying are unattended computers with no screensaver passwords.

Successful social engineers make an attack more effective by using technology instead of just using the telephone or people's weaknesses. Password-cracking software, spy software recording all keyboard movements and software used to prepare files containing harmful codes may be given as examples of this effort.

A successful social engineer mostly targets employees at the lower layers of the intra-organizational power ranking, with particular emphasis on the fresh recruits. But why fresh recruits? This question may be replied saying that most of the time they are unaware about the potential outcomes of organizational information or some of their actions and that they can easily be put under influence.

Figure 1 shows the social engineering attack process. The first stage of the process is the research process. At this stage, information is obtained as much as possible about the person or organization to be attacked. Then is the stage of arousing friendship and trust. The social engineer tries to gain the sympathy of the victim using the information it has obtained during the previous stage. The next stage starts if he can be successful and the trust is misused. Sensitive information obtained is then evaluated. If this information is adequate, then the attack is finalized. If the obtained information is inadequate, the attacker goes back to the research stage and the cycle is repeated once again by the social engineer.

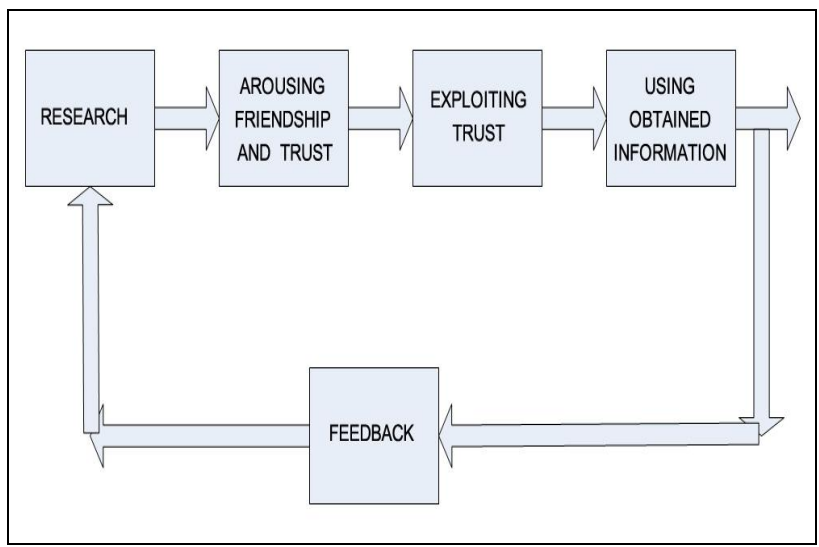

Figure 1. Social engineering cycle

## 2.2 PROTECTION METHODS

User awareness trainings containing the following articles should be made to all agency personnel in certain periods [2], [3], [5], [9], [10], [19], [21]:

- Importance of procedures and their applications,
- Login to the computer and password security,
- Making hardware and software changes in the computer,
- Use of laptops,
- File access and sharing,
- Use of printers,
- Use of portable media,
- Virus protection,
- Internet access security,
- E-mail security,
- Backing up,
- Computer security events notification,
- Social engineering.

Within the scope of the "Continued Awareness Program", measures like posting caricatures and hints, posting the photos of month's security personnel, posting announcements on bulletin boards, posting various information security posters, sending memo emails, following information security-related internet sites, distributing brochures and using security-related screensavers and background pictures should be taken in the intranet of the organization for information security.

Besides, risk analysis studies should be made in the organization. In this regard, the agency's informational assets, potential threats to these assets and potential damages that these threats might cause need to be determined. Data classification studies should be made at the organization.

Following measures would be useful aside from the above-mentioned measures:

- Personnel identification cards should be worn by all personnel.

- An information security division and event notification center should be established within the organization.
- Periodical Information Security Tests should be made in the organization.
- Antivirus software must be installed on all computers and the definition file must be kept update.
- Documents important for the agency that need to be thrown in the waste bin should be destroyed by paper clippers.
- Password-protected screensavers should be used on computers.
- Procedures should be prepared containing the directives that should be observed by quitting personnel.
- ID cards of visitors should be collected and a member of the Agency staff should accompany this visitor.
- Strong passwords are must for the organization and these passwords should never be written on anywhere or shared with others.

## 3. RESULTS AND DISCUSSION

TUBITAK UEKAE Department of Information Systems Security makes social engineering attacks to public agencies for about 3 years. To this end, this test was so far conducted on 6 public agencies. Those agencies are the leading agencies in monetary and finance, health, telecommunications, and applied sciences in Turkey; so capturing sensitive information from those agencies which are the main parts of the critical infrastructure of Turkey is crucial. Within this scope, telephone conversations were made with 56 users and passwords of 38 of them (approximately 68% of users) were acquired.

Figure 2 shows the methods used at the research stage of the social engineering cycle. Web sites of the agencies were analyzed to obtain information about the organizations, with an effort to find information about the staff and intra-organizational instructions and procedures. Information was also obtained about organizations offering assistance to those agencies, if any.

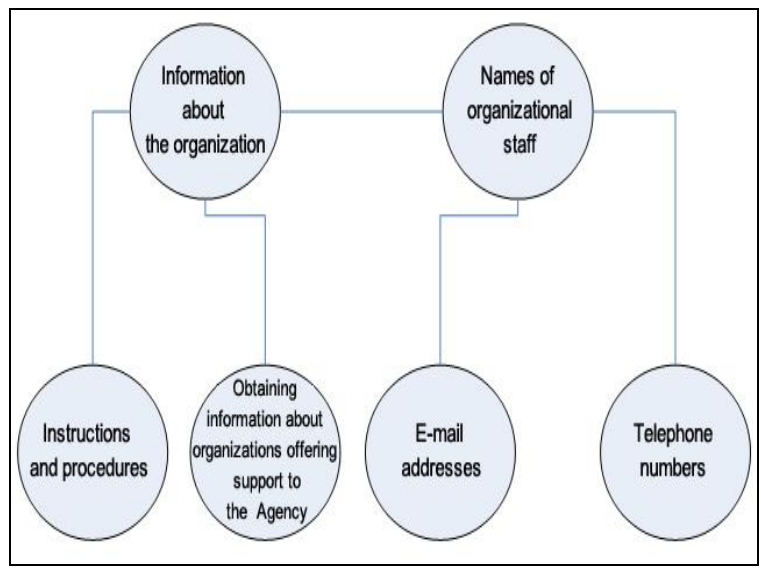

Figure 2. Methods used during research stage

All these information is brought together and telephone conversations were made with the agency's staff to obtain sensitive information (login passwords for computers used by agency staff or software passwords). The social engineer has introduced himself as a new employee in IT department. After the social engineering test was finished, the IT personnel was informed to ask users whose sensitive information were acquired during the test to change their passwords.

In Agency A, 10 users were interviewed and sensitive information of 8 of them were obtained. And in Agency B and E, 10 users were interviewed and sensitive information of 5 of them were obtained. In Agency C and D, 10 and 8 users were interviewed respectively and sensitive information of 6 each of them were obtained. Finally, in Agency F, 8 users were interviewed and sensitive information of 8 of them were obtained (Figure 3). While the success rate (number of password-obtained participants/total number of participants) in Agency A was 80%, this rate was 50% for Agency B and E, 60% in Agency C and 75% in Agency D, and 100% in Agency F which are seriously high rates. In fact, it may not be quite true to consider success rate by Agency because even if only one person's information is obtained from the organization as the attacker, it may not be possible to be able to reach this information with this sensitive information. When the results are analyzed, unfortunately it is apparent that civil servants are poorly aware of information security.

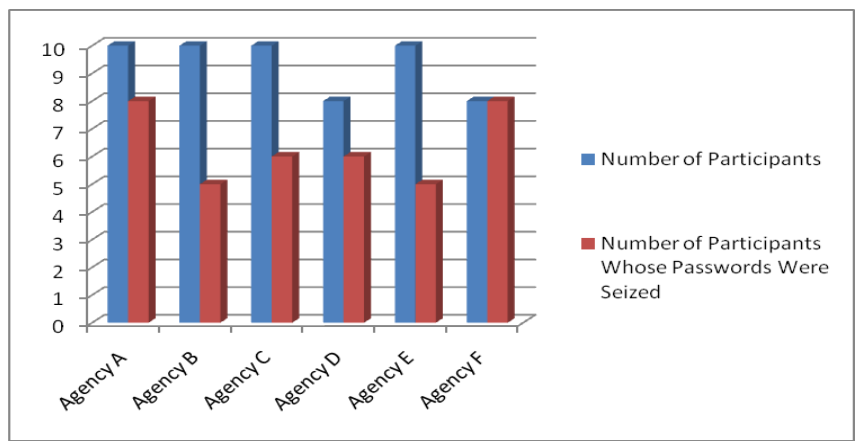

Figure 3. Results of social engineering in Turkish public agencies

## 4. CONCLUSIONS

This paper has made an analysis of the social engineering tests that have been carried out in several Turkish public agencies. The principal contribution of the paper is to calculate the percentage of employees in Turkish public agencies who have shared their sensitive information which should never be shared by anyone else, and determine the reasons of sharing sensitive information by stating fallacies and related amendments through interviews by employees so as to understand the user behavior as regard as a social engineering attack is carried out. The aim of the research is to figure that the employees in Turkish public agencies have a lack of information security awareness and they compromise the information security principles which should be necessarily applied for any public agencies.

Now, information security professionals should recognize that the organization's information security should be more important than computer, software, or hardware security [13]. The threat might come from anywhere and at any time. Staff members should be precautious about requests from unknown people and personal information (e.g., passwords) should not be shared with anyone including system managers, coworkers, and even managers. Moreover, information security awareness trainings should be periodically given to all personnel in the organization. Finally, conducting periodical Information Security Tests including social engineering attack tests in the organization would be useful.